\documentstyle[11pt,appbx2,epsfig]{article}
\begin{document}
\title{PROTON INDUCED $K^+$ PRODUCTION AND \\
THE $s\bar{s}$ CONTENT OF THE PROTON
\thanks{Presented at the "Meson 96" Workshop, Cracow, Poland, May 10 -- 
14,
1996}}
\author{F.\ Kleefeld${}^{\mbox{\rm \scriptsize a}}$, M.\ Dillig${}^{\mbox{\rm
\scriptsize a}}$ and F.\ Pilotto${}^{\mbox{\rm \scriptsize a,b}}$ 
\address{
${}^{\mbox{\rm \scriptsize a}}$ Institute for Theoretical Physics III, University of Erlangen-N\"urnberg\\
D-91058 Erlangen, Germany\\
${}^{\mbox{\rm \scriptsize b}}$ Instituto de F\'\i sica, UFRGS, Porto Alegre, RS, Brazil}
}
\setcounter{page}{2867}
\maketitle

\begin{abstract}
We investigate the exclusive proton--induced $K^+$
 production near the kaon threshold. Compared are two models: a meson--exchange
 model, which includes $\pi , \rho , K$ and $K^\ast$  exchange, together with
 the
 dominant baryon resonances, and a quark--gluon model, where momentum
 sharing is provided via the exchange of two gluons. Based on covariant bound
 state wave functions for the meson and the baryons --- which we treat as
 quark--diquark objects --- first results for the $pp\rightarrow p\Lambda K^+$
 total cross section are presented for both models. \\[5mm]
\end{abstract}

     Numerous calculations in different QCD--inspired quark models have
 indicated that mesons and baryons can be described qualitatively as
 objects, which involve dominantly $q\bar{q}$ and $3q$ valence quarks,
 respectively.
 Recently, however, particularly with the so called "spin crisis of the
 nucleon", the question on the admixture of seaquarks in hadrons has been
 raised \cite{Al}. In view of recent experimental activities at various hadron factories,
 such as COSY or CELSIUS, one promising source of information on the $s\bar{s}$
 content of the proton is the exclusive $K^+$ production in $pp\rightarrow
 NYK$, with $Y$ being a $\Lambda$ or $\Sigma$ hyperon: in the simplest
 picture, the production process can be viewed as the excitation of an $s\bar{s}$
 pair in the proton, which is subsequently --- during the scattering process ---
 put on its mass shell.

 The standard approach to meson production is the meson--exchange model, where
 the large momentum transfer of typically $\ge$ 1 GeV/c is shared between the
 hadrons by meson exchange \cite{Br}\cite{La} (Fig.\ 1a).
 %========Figure ==========================================================
 \begin{figure}[htbp]
 \epsfxsize=  110.0mm
 \epsfysize=  5.0cm
 \centerline{\epsffile{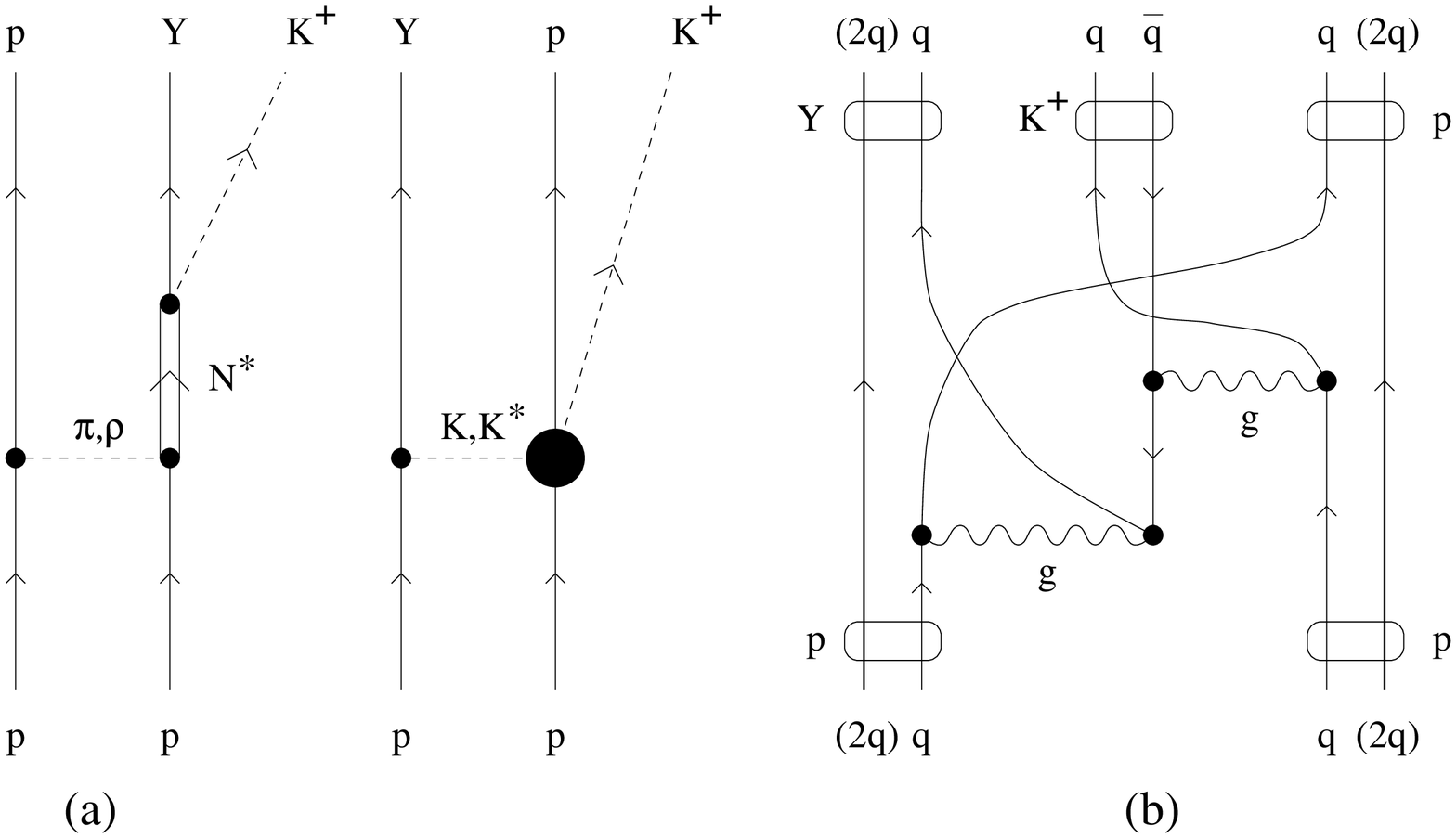}}
 \caption{Meson--exchange model (a) and quark--gluon model (b) for
 $pp$--induced $K^+$
 threshold production} 
 \end{figure}
 %=======End of Figure =====================================================
 The dominant contributions are expected from the exchange of $(\pi , \rho)$ in
 the nonstrange and $(K,K^\ast )$ in the strange meson sector, whereas the
 rescattering amplitude \newpage
 \noindent is saturated by s-- and p--wave baryon--resonances
 in the region of 1.6 to 1.8 GeV. 
 Preliminary results show that the model reproduces the scarce existing data
\cite{Fl} in a qualitative form. However, the approach shows serious shortcomings, resulting from
 its --- basically --- nonrelativistic nature and the significant sensitivity of the
 results on the parametrization of off--shell effects in the various
 meson--baryon vertices.

 The intrinsic short range scale $\lambda \sim 0.2-0.5$ fm of proton induced $K$
 production suggest a parametrization of the production mechanism in terms of
 quark and gluon degrees of freedom. Guided by experience from exclusive high momentum
 transfer reactions, we employ a simple gluon exchange model: we
 assume that the large momentum is
 shared by a two--gluon--exchange mechanism between the interacting constituent quarks, followed by the appropriate
 interchange of quark lines (to guarantee colourless objects with the right
 flavour content in the final state; compare Fig.\ 1b). Soft corrections of
 higher order are included in addition as standard initial and final state
 corrections.

 We briefly summarize the basic ingredients. 
 To allow for an appropriate organization of the Fock components, we work in
 the instant (light cone) form.
 For a covariant description we
 start from the Bethe--Salpeter equation,
 \begin{equation} \Psi (P,p) \quad = \quad G_{BS}(P,p) \int K(P,p,k) \,\,\Psi (P,k) \,dk
 \end{equation}
 which is reduced to a 3--dimensional covariant Quasi--Potential equation upon
 integrating out the relative energy component in the BS--Propagator \newpage
 \noindent in the
 spirit of a spectator picture for the bound particles \cite{Gr},
 schematically
 \begin{equation} G_{BS}(P,p) = \frac{1}{(D_1(P,p)+i\varepsilon )
 (D_2(P,p)+i\varepsilon )} \equiv
 - \, \frac{i\pi \delta (D_1(P,p) - D_2(P,p))}{D_1(P,p)} 
 \end{equation}
 which amounts for the transisition to the light cone in the substitution
 \begin{equation}
 P = (P_0,P,\,\b{0}) \Rightarrow (E,1,\b{0}) (\mbox{we used } P_+=1); 
 p = (p_0,\b{p})   \Rightarrow (E\,x,\b{p}_\perp ),
 \end{equation} 
 where $E$ is the total energy of the bound hadron and $x$ the momentum
 fraction carried by the constituent. The interaction kernel is approximated by a
 one--gluon exchange potential 
 \begin{equation} K(P,p,k) \Rightarrow 
 K_{OBE} (x,\b{p}_\perp ) =  
 \frac{C}{(E\,x)^2 + \b{p}_\perp^2 }\,\,\cdot \,\,
 \frac{\alpha_s (p_0^2)}{\displaystyle \ln \left( 1 + 
 \frac{(E\,x)^2 + \b{p}_\perp^2}{\Lambda_{QCD}^2}\right)} 
 \end{equation}
 ($C$ absorbs spin and colour factors).
% The resulting configuration space one--gluon exchange potential combines in the
% heavy mass limit $E\rightarrow\infty$
% \begin{equation} V_{OGE} (r) \quad = \quad \mbox{const}
% \,\alpha_s (p_0^2) \left( - \,\frac{1}{r} + \Lambda_{QCD}^2 \,r + 
% \frac{e^{-\mu r}}{r} \left( \sin\mu r + \cos\mu r\right)\right) 
% \end{equation}
% both confinement and asymptotic freedom. 
As a final step we describe the baryons
 as Quark--Diquark objects and parametrize their wavefunctions as
 \begin{equation} \Psi (x,\b{p}_\perp ) \quad = \quad 
 N \,\,{\left(\prod\limits_{i=1}^3 \left(\b{p}_\perp^2 + 
 \alpha_i^2 (x)\right)\right)}^{-1}
 \end{equation}
\begin{eqnarray} \alpha_1^2 (x) & = & 
 (\eta +x) \left( M^2 - m^2 - (1-\eta -x) \, E^2 \right) + m^2 \nonumber \\ 
  \alpha_{2,3}^2 (x) & = &
 x\, \left( (x + 2\eta -1) E^2 +M^2 -m^2 \right) + \Lambda_{2,3}^2 
 \end{eqnarray}
 with the quark and diquark masses $m$ and $M=2m$, $\eta = m/(m+M)$, and the scale
 parameters $\Lambda_2=\Lambda_q$ and $\Lambda_3=\Lambda_{qq}$.
 In the spirit of the impulse approximation the strangeness content in the proton is
 defined in the model via the relation
 \begin{eqnarray} & &\Psi_{3q,s\bar{s}} (P,\b{q},\b{k}) = 
 G_{3q,s\bar{s}} (P,\b{q},\b{k}) \nonumber \\
 & & \ast \int d\b{p} \left\{ 
 V_{q\rightarrow qs\bar{s}} (P,\b{q},\b{k};\b{p}) +
 V_{2q\rightarrow 2q,s\bar{s}} (P,\b{q},\b{k};\b{p}) \right\}
 \, \Psi_{3q} (P,\b{p})
 \end{eqnarray}
 (the notation is understood as light cone coordinates).
Preliminary results for the two--gluon--exchange model are shown in Fig.\ 2b (for
$m_{u,d}=330$ MeV, $m_s=505$ MeV, $\Lambda_{QCD}=250$ MeV and 
$\Lambda=\Lambda_q=\Lambda_{qq}$).
In spite of the qualitative agreement with the data, a detailed conclusion on
the $s\bar{s}$ content of the proton requires a further study of the numerics
and a detailed test of the consistency of the model for other meson channels.
Work\newpage
\noindent along this line is in progress \cite{Kl}.
 This work was supported in part by the Kernforschungszentrum
 J\"ulich under contract No. ER-41154523.
 %========Figure ==========================================================
 \begin{figure}[t]
 \epsfxsize=  126.0mm
 \epsfysize=  5.5cm
 \centerline{\epsffile{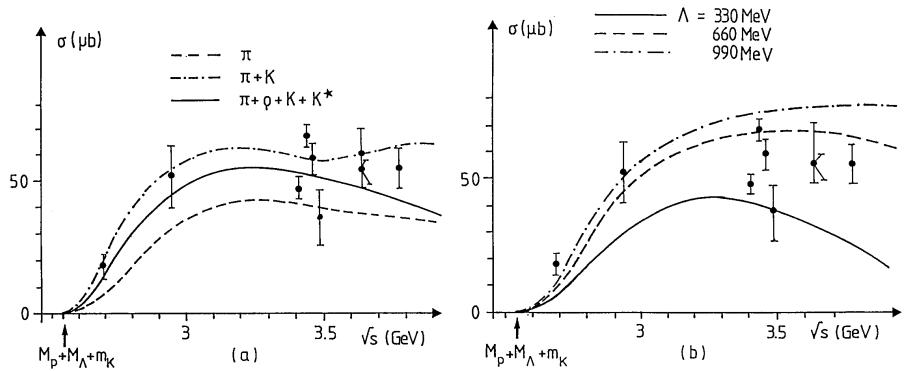}}
 \caption{Preliminary results for the energy dependence of
 $\sigma_{pp\rightarrow p\Lambda K^+}$ in the (a) meson--exchange and (b) two
 gluon--exchange model (comp.\ Fig. 1).}
 \end{figure}
 %=======End of Figure =====================================================
 %======================
 %\section*{References}
 %======================


\begin{thebibliography}{99}
\bibitem{Al} M.\ Alberg, J.\ Ellis, D.\ Kharzeev, 
               {\it Phys.Lett.,} 113 {\bf B356} (1995). 
\bibitem{Br} G.\ Brown {\it et al.,} 
               {\it Phys.Rev.} {\bf C43} 1881 (1991). 
\bibitem{La} J.M.\ Laget, 
               Internal report DAPNIA/SPhN 9250 (Saclay, 1992). 
\bibitem{Fl} V.\ Flaminio {\it et al.,} Compilation of cross sections;
III: $p$ and $\bar{p}$ induced reactions; CERN-HERA 84-01 (Genf, 1984). 
\bibitem{Gr} F.\ Gross, 
               {\it Relativistic Quantum Mechanics and Field Theory,}
                 J.Wiley \& Sons, Inc., New York, 1993.                  
\bibitem{Kl} F.\ Kleefeld, 
               Thesis, Univ. Erlangen (in preparation). 
 \end{thebibliography}
 \end{document}